# Reduced Thermal Conductivity of Supported and Encased Monolayer and Bilayer MoS$_2$

Alexander J. Gabourie[1], Saurabh V. Suryavanshi[1], Amir Barati Farimani[2], and Eric Pop[1,3,4,*]

[1]Department of Electrical Engineering, Stanford University, Stanford, CA 94305, USA

[2]Department of Mechanical Engineering, Carnegie Mellon University, Pittsburgh, PA 15213, USA

[3]Department of Materials Science & Engineering, Stanford University, Stanford, CA 94305, USA

[4]Precourt Institute for Energy, Stanford University, CA 94305, USA

*Contact: *epop@stanford.edu*

**Abstract:**

Electrical and thermal properties of atomically thin two-dimensional (2D) materials are affected by their environment, e.g. through remote phonon scattering or dielectric screening. However, while it is known that mobility and thermal conductivity (TC) of graphene are reduced on a substrate, these effects are much less explored in 2D semiconductors such as MoS$_2$. Here, we use molecular dynamics to understand TC changes in monolayer (1L) and bilayer (2L) MoS$_2$ by comparing suspended, supported, and encased structures. The TC of monolayer MoS$_2$ is reduced from ~117 Wm$^{-1}$K$^{-1}$ when suspended, to ~31 Wm$^{-1}$K$^{-1}$ when supported by SiO$_2$, at 300 K. Encasing 1L MoS$_2$ in SiO$_2$ further reduces its TC down to ~22 Wm$^{-1}$K$^{-1}$. In contrast, the TC of 2L MoS$_2$ is not as drastically reduced, being >50% higher than 1L both when supported and encased. These effects are due to phonon scattering with remote vibrational modes of the substrate, which are partly screened in 2L MoS$_2$. We also examine the TC of 1L MoS$_2$ across a wide range of temperatures (300 to 700 K) and defect densities (up to $5\times10^{13}$ cm$^{-2}$), finding that the substrate reduces the dependence of TC on these factors. Taken together, these are important findings for all applications which will use 2D semiconductors supported or encased by insulators, instead of freely suspended.

I. INTRODUCTION

Two-dimensional (2D) semiconducting $MoS_2$ is a promising material for technologies beyond silicon [1, 2], flexible and transparent electronics [3, 4], and thermoelectric applications [5, 6]. However, it is known that electrical and thermal conductivities in other atomically thin 2D materials, like graphene, degrade when in contact with a substrate due to scattering with substrate impurities or remote phonons [7-9]. This occurs because electron and phonon wavelengths are comparable to or larger than the 2D material thickness, especially in monolayers. In realistic applications, 2D materials will almost always be used in contact with a metal (for contacts) or an insulator (for gate dielectrics, substrates, or encapsulation layers), thus it is important to understand their thermal properties in this context. Moreover, it is already known that thermal bottlenecks limit nanoelectronics performance with traditional semiconductors [10, 11]. Therefore, such considerations must be included when evaluating future 2D material applications.

Thermal transport in 2D materials is fundamentally different than in bulk because transport is confined to two dimensions and 2D material interfaces are dominated by van der Waals (vdW) interactions, which present a bottleneck to heat removal from 2D devices [12, 13]. In fact, the $MoS_2$-$SiO_2$ vdW interface is known to have a large thermal resistance, equivalent to that of ~90 nm of $SiO_2$ [12]. To understand its effect in the context of a 2D device, we refer to Fig. 1. For a thermally "long" 2D device (of length $L \gg 3L_H$, where $L_H \sim 0.1$ μm is the thermal healing length [14]), the temperature rise is mostly determined by this interfacial thermal resistance [14, 15]; however, for thermally "short" devices ($L < 3L_H$) heat can be removed at the contacts and the temperature rise strongly depends on the thermal conductivity of the 2D material [15]. Given that 2D devices have already been demonstrated with sub-100 nm dimensions [16-19], it is crucial to determine how the substrate limits the thermal conductivity of 2D materials like $MoS_2$.

Thermal measurements of geological, *bulk* $MoS_2$ samples have estimated an in-plane thermal conductivity [20-24] of 82 to 110 $Wm^{-1}K^{-1}$ and cross-plane thermal conductivity of ~5 $Wm^{-1}K^{-1}$ [25]. On the other hand, a vast majority of measurements of monolayer (1L) $MoS_2$ exist only for freely *suspended* samples, revealing a range of 13 to 97 $Wm^{-1}K^{-1}$ for the in-plane thermal conductivity, ostensibly due to sample-to-sample variation between mechanically exfoliated [26, 27] and chemically synthesized samples [28-32]. (Below, we will see that variation in defect densities could explain this measurement variation.) Simulation efforts also display large variations due to different techniques or inter-atomic potentials and have only focused on bulk or suspended $MoS_2$. Interestingly, most simulations find the 1L in-plane thermal conductivity to fall in two distinct ranges at 300 K (excluding studies with extremely high or low values), i.e. between 19 to 38 $Wm^{-1}K^{-1}$ on the low end [26, 33-41] and between 82 to 178 $Wm^{-1}K^{-1}$ on the high end [30, 42-56], with no calculations falling between the two (as summarized in Supplementary Fig. S1).

Despite these efforts, there are no simulation studies on the thermal conductivity of MoS$_2$ *supported* or *encased* by an insulator such as SiO$_2$, and only one recent experiment [57]. In contrast, several studies of substrate-supported [7, 8, 57-59] or encased [60] graphene have found its thermal conductivity reduced by 5-10× or 30-40×, respectively, compared to suspended graphene, agreeing well with simulations [8, 61]. Similarly, simulations of SiO$_2$-supported silicene [62] predict ~78% thermal conductivity degradation compared to suspended silicene. One reason for the lack of thermal measurements on other supported or encased 2D semiconductors is that their thermal conductivity is much lower than graphene. This makes it difficult to distinguish the heat flowing laterally in the ultrathin 2D material vs. the much thicker supporting or encasing insulator. Thus, this is an area where atomistic simulations can play an important role, not only to quantify the effects of the adjacent insulator on the thermal conductivity of a 2D semiconductor, but also to provide physical insight into why this occurs, opening the door for tuning such 2D material properties. This is the aim of the present study, with respect to 1L and 2L MoS$_2$, supported and encased by SiO$_2$.

## II. SIMULATION METHODOLOGY

The thermal conductivities of crystalline materials can be well described by the Peierls-Boltzmann transport equation paired with calculations from density functional theory [63]. However, when systems have broken symmetry, these methods struggle to reproduce experimental measurements [64]. Such is the case for supported 2D material systems (which include interfaces) and disordered materials (i.e. amorphous, defective), both breaking symmetry. Given that adequate material potentials exist, classical molecular dynamics (MD) can overcome these limitations and accurately model all anharmonicities and phonon-phonon interactions at relevant length and time scales [51]. Recently, a comparative study of empirical MD potentials has determined the optimal potential for thermal transport in MoS$_2$ [51], making MD a more attractive method. For these reasons, we choose to use MD for all simulations in this study.

All results in this study are calculated using the Graphics Processing Units Molecular Dynamics (GPUMD, here version 2.1) package [65-67]. For supported and encased MoS$_2$ calculations, we modified the GPUMD package to isolate only the MoS$_2$ contributions to thermal conductivity. We use the LAMMPS package [68, 69] to check for consistent forces between the different simulation packages. To model the atomic interactions in MoS$_2$, as well as between layers of MoS$_2$, we use the reactive empirical bond-order potential with a Lennard-Jones addition (REBO-LJ) [70, 71]. The REBO-LJ implementation in GPUMD has a modification introduced by Stewart and Spearot [72], and we use the LJ parameters designed for a 300 K crystal temperature [71]. We model the SiO$_2$ with the Tersoff potential [73] parameterized by Munetoh *et al.* [74] and the MoS$_2$-SiO$_2$ van der Waals (vdW) interactions with the LJ potential using the Lorentz-Berthelot mixing rules. The mixing parameters are listed in Table S1 of the Supplement.

All simulations are based on three structures: suspended MoS$_2$, MoS$_2$ supported on amorphous SiO$_2$ (a-SiO$_2$), or MoS$_2$ encased (top and bottom) by a-SiO$_2$. We study both monolayer and bilayer MoS$_2$ in these scenarios. The simulation cell areas are 10×10 nm$^2$, the MoS$_2$ monolayer is 6.15 Å thick, consistent with experimental observations [75], and the a-SiO$_2$ substrate is 5.4 nm thick as shown in Fig. 1(c). We chose our simulation cell area to be the minimum size needed to reproduce the 1L MoS$_2$ thermal conductivity results from previous work [51] and found the MoS$_2$ thermal conductivity to be independent of a-SiO$_2$ thickness (see section 3 in the Supplement). We use periodic in-plane boundary conditions (BCs) to model an infinite MoS$_2$ sheet and minimize finite-size effects on phonon mean free paths. In the out-of-plane direction, vacuum and free BCs are used. The a-SiO$_2$ is created by a simulated annealing, the details of which can be found in section 4 of the Supplement.

To compute the thermal conductivity we use the homogenous nonequilibrium MD (HNEMD) method [76]. This method is consistent with, but more efficient than, the commonly used equilibrium MD (EMD) and nonequilibrium MD (NEMD) methods [77], and it does not have boundary scattering because of periodic BCs in the transport direction. The HNEMD method requires an additional driving force parameter $\boldsymbol{F}_e$ to calculate an applied external force [76]. Because MoS$_2$ has hexagonal symmetry, the intrinsic in-plane thermal conductivity is isotropic. As such, we apply the driving force parameter in only one direction, reducing it to a scalar. Here we choose $F_e$ = 0.2 μm$^{-1}$, consistent with previous HNEMD simulations for MoS$_2$ [51]. The thermal conductivity $\kappa$, with a simplification due to isotropy, is then [51, 76]:

$$\kappa(t) = \frac{1}{t}\int_0^t \frac{\langle J(\tau) \rangle_{\mathrm{ne}}}{TVF_e} d\tau \tag{1}$$

where $J$ is the heat current, $T$ is the temperature, and $V$ is the system volume. The integral represents a post-processed, running average of thermal conductivity over a simulation up to time $t$. The integrand is the direct thermal conductivity calculated by GPUMD. Using a time step of 0.5 fs, we output average heat current every 500 fs, and use Eq. (1) to compute the final value of thermal conductivity, which converges by 10 ns. (Additional simulation details are given in Supplementary section 5.)

Due to the influence of the driving force parameter ($F_e$) on the heat flux and the direct calculation of thermal conductivity, the HNEMD method is able to compute the substrate-supported thermal conductivity, a situation where using the EMD method was shown to be challenging [62]. Furthermore, GPUMD decomposes the in-plane thermal conductivity into contributions from in-plane atomic motion (dominant in longitudinal and $x$-$y$ transverse phonons) and out-of-plane atomic motion (dominant in flexural phonons) [52]. Schematics of phonons related to each type of motion are shown [78] in Fig. 1(d) and further discussed below. More details about the HNEMD thermal conductivity and the GPUMD heat flux formulation can be found in Refs. [76] and [79]. The final thermal conductivity of each simulation is taken to be $\kappa(t = 10$ ns) using Eq. (1). Our reported values are averaged over $n = 10$ independent runs (i.e. simulations with different initial velocities) with a standard error of $\sigma/\sqrt{n}$, where $\sigma$ is the standard deviation of $\kappa(t = 10$ ns) values over the $n$ independent runs.

### III. RESULTS AND DISCUSSIONS
#### A. Monolayer MoS$_2$

We first calculate the in-plane thermal conductivity of suspended and SiO$_2$-supported monolayer MoS$_2$ in Figs. 2(a) and 2(b) respectively, including its decomposition into in-plane and out-of-plane atomic motion contributions. The *suspended* 1L MoS$_2$ thermal conductivity (converged at $t = 10$ ns and averaged over $n = 10$ independent runs) is $\kappa = 117.0 \pm 2.0$ Wm$^{-1}$K$^{-1}$ (in agreement with measurements of bulk [20] MoS$_2$ and recent simulations [51]) with contributions from in-plane atomic motion of $85.9 \pm 2.1$ Wm$^{-1}$K$^{-1}$ and out-of-plane motion of $31.1 \pm 1.6$ Wm$^{-1}$K$^{-1}$. In contrast, we find the in-plane thermal conductivity of MoS$_2$ *supported* on a-SiO$_2$ to be $30.9 \pm 1.5$ Wm$^{-1}$K$^{-1}$ (~74% decrease) with in-plane and out-of-plane contributions of $26.3 \pm 1.2$ Wm$^{-1}$K$^{-1}$ (~69% decrease) and $4.6 \pm 0.7$ Wm$^{-1}$K$^{-1}$ (~85% decrease), respectively. We note this result is smaller than the $63 \pm 22$ Wm$^{-1}$K$^{-1}$ recently measured for SiO$_2$-supported MoS$_2$ [57]; however, a sputtered, 20 nm Ni capping layer may have affected these in-plane thermal conductivity measurements.

While our simulations show a greater proportion of the out-of-plane contribution is damped on a-SiO$_2$, the reduction of the in-plane contribution drives the overall reduction in thermal conductivity. This contrasts the thermal conductivity reduction in supported graphene, which, experimentally and through simulation, has been shown to suffer an ~80% to ~90% degradation mostly from the damping of its out-of-plane motion (which directly corresponds to flexural phonons in graphene) [7, 8, 61]. The difference is due to the dominant mode of thermal transport in MoS$_2$ and graphene. Graphene follows a symmetry-based selection rule that restricts anharmonic phonon-phonon scattering of flexural modes [80] leading to an *out-of-plane* contribution that carries approximately 2× more heat than in-plane [81]. Monolayer MoS$_2$ is three atoms thick and does not follow this rule, leading to an *in-plane* contribution that carries more than 2× the heat of its out-of-plane contribution. Thus, our findings show that the suppression of the *dominant* mode of thermal transport (out-of-plane atomic motion for graphene, in-plane for MoS$_2$) drives the overall reduction of thermal conductivity in supported 2D materials, not *only* the dampening of the out-of-plane motion.

To better understand thermal transport in supported MoS$_2$, we plot the vibrational density of states (VDOS) of both suspended and supported MoS$_2$ with a-SiO$_2$ in Figs. 2(c) and 2(d), respectively. From either VDOS plot, we can see that molybdenum contributions to the overall VDOS are much larger than sulfur below 8 THz. This is the frequency range of the acoustic modes which are the main heat carriers [56], meaning that much of the thermal transport is carried out by vibrations of molybdenum atoms. Because the acoustic modes of MoS$_2$ do not appear to be affected by the substrate, we conclude that additional *scattering* with the SiO$_2$ causes the reduction in thermal conductivity of supported MoS$_2$. This is confirmed by Fig. 2(d) which reveals a significant overlap of the a-SiO$_2$ and MoS$_2$ VDOS, especially at the lower frequencies of the heat-carrying acoustic modes. The supported MoS$_2$ phonons have substantially more modes (including substrate vibrations) to interact with, i.e. through anharmonic scattering or harmonic energy transfer [82], disrupting thermal transport in the MoS$_2$ and reducing its thermal conductivity. This phenomenon is similar to that of remote phonon scattering for the reduction of transistor mobility in ultrathin films or silicon inversion layers [83, 84]. For additional details on the calculation of the VDOS, see Supplementary section 7.

**B. Temperature Dependence**

MoS$_2$ electronic devices will sometimes operate several hundred Kelvin above room temperature as seen already in self-heating field-effect transistors [85], and as desired for some thermoelectric applications [86]. Thus, we also investigate the thermal conductivity of both suspended and supported MoS$_2$ from 300 K to 700 K. This range is above the Debye temperature ($\theta_D$) of MoS$_2$, ensuring the validity of these classical MD simulations without the need for quantum corrections [87]. For suspended 1L MoS$_2$, previous calculations [38] have placed the Debye temperature at $\theta_D \approx 262$ K, with bulk MoS$_2$ measured to be $\theta_D \approx$ 260-320 K [88].

The simulation results for suspended 1L MoS$_2$ are shown in Fig. 3(a), revealing a steep temperature-related decline for both in-plane and out-of-plane atomic motion contributions. The overall reduction of thermal conductivity with temperature scales as $\kappa \propto T^{-1.94}$ (solid black line) which implies a stronger contribution of four-phonon scattering [89] ($\kappa \propto T^{-2}$; more common at high $T \gg \theta_D$ which also plays a role at high temperature in Si and Ge) [90, 91] than of three-phonon Umklapp scattering ($\kappa \propto T^{-1}$; dashed black line).

This temperature dependence appears stronger than in other suspended low-dimensional materials, such as carbon nanotubes and graphene (with natural concentrations of $^{13}$C isotopes), which experimentally show a $T^{-x}$ dependence, with $1.1 < x < 1.3$ [92, 93]. However, the carbon nanotube and graphene data do not probe temperatures above their $\theta_D$, which is very high, $\theta_D \approx 2100$ K [94]. In addition, the temperature dependence of isotopically pure graphene (0.01% $^{13}$C) was found to be steeper than for natural graphene [93]. Since our modeled Mo and S masses are weighted averages over naturally occurring isotopes, we effectively have an isotopically pure system (i.e. one mass for each atom type), which may explain why the temperature dependence we find here for MoS$_2$ is similar to isotopically pure graphene. Relevant details on the kinetic theory and fitting can be found in Supplementary sections 8 and 9, respectively.

Figure 3(b) shows the temperature dependence of *supported* 1L MoS$_2$ is substantially different than the suspended 1L MoS$_2$. We note that, since the out-of-plane motion is already severely damped by the substrate, the in-plane contribution dominates the total thermal conductivity reduction with temperature. Comparing our calculations to kinetic theory again, we find the temperature decay scales as $\sim T^{-1.2}$, which suggests that three-phonon processes dominate the reduction of thermal conductivity with temperature. However, this is not necessarily an accurate characterization of the dominant phonon processes in MoS$_2$ because we cannot decouple its intrinsic scattering events from those involving the substrate vibrations. We do know that the four-phonon processes, which are influential in our suspended MoS$_2$, are overwhelmed by the effects of the substrate.

## C. Defect Dependence

It is known that the properties of 2D materials are degraded or altered by defects that are either naturally occurring or introduced during growth or layer transfer processes [95]. Here, we study the effects of the most common defect type, zero-dimensional sulfur vacancies [95-99], on the thermal conductivity of MoS$_2$. Previous experimental studies have reported sulfur vacancy densities from $n_v = 7 \times 10^{10}$ cm$^{-2}$ to $6.5 \times 10^{13}$ cm$^{-2}$ for exfoliated MoS$_2$ or MoS$_2$ grown by chemical vapor deposition (CVD) [97-101]. For this set of simulations, we randomly introduce sulfur vacancies such that their density ranges from $10^{12}$ cm$^{-2}$ to $5 \times 10^{13}$ cm$^{-2}$, which corresponds to 1 to 50 sulfur vacancies in the simulated $10 \times 10$ nm$^2$ MoS$_2$ sheet. Overall, there are no vacancy clusters and we expect similar trends for different single sulfur vacancy configurations [54].

The results for *suspended* MoS$_2$ are shown in Fig. 4(a), with vacancy-free calculations plotted left of the *x*-axis break for reference. Here we find that a small vacancy density of $10^{12}$ cm$^{-2}$ already reduces the total thermal conductivity by ~19%. For the vacancy densities studied, the calculated thermal conductivity range is $94.4 \pm 3.1$ Wm$^{-1}$K$^{-1}$ to $30.7 \pm 2.5$ Wm$^{-1}$K$^{-1}$, which encompasses the experimental results of ~84 Wm$^{-1}$K$^{-1}$ for exfoliated 1L MoS$_2$ and ~30 Wm$^{-1}$K$^{-1}$ for CVD-grown 1L MoS$_2$ [27, 29]. This relationship between our calculations and experiment is not unexpected because CVD-grown MoS$_2$ could be more defective than exfoliated [97, 98], particularly at the time of the measurements referenced here. Recent Peierls-Boltzmann transport calculations have also pointed to defects when explaining the large range in reported experimental thermal conductivities [54].

We also find that the contribution from out-of-plane motion is less sensitive to defects than the in-plane contribution. This imbalance is most severe at our lowest defect density ($10^{12}$ cm$^{-2}$) as the out-of-plane contribution, when compared to our defect-free structure, is reduced by only ~7% compared to the

~24% reduction of the in-plane contribution. The dashed line in Fig. 4(a) plots the expected defect dependence trend based on kinetic theory ($\kappa \propto \sim n_v^{-1}$) [89]. The suspended MoS$_2$ follows this trend with small deviations at extremes. More information on the relevant kinetic theory and fitting can be found in Supplementary sections 8 and 9, respectively.

Compared to defective suspended MoS$_2$, defective *supported* MoS$_2$ has a thermal conductivity that is less sensitive to changes in vacancy density. Figure 4(b) reveals that the thermal conductivity only decreases by ~5.5% at a vacancy density of $10^{12}$ cm$^{-2}$ compared to defect-free, supported MoS$_2$. At the highest vacancy density, $5\times10^{13}$ cm$^{-2}$, the thermal conductivity has decreased by ~58% to $12.9 \pm 1.2$ Wm$^{-1}$K$^{-1}$, which comes from a ~62% decrease to $10.0 \pm 1.0$ Wm$^{-1}$K$^{-1}$ from the in-plane contribution and ~36% decrease to $2.9 \pm 0.6$ Wm$^{-1}$K$^{-1}$ from the out-of-plane contribution, with respect to defect-free, supported MoS$_2$. We note that the out-of-plane contribution is not as sensitive to defects as the in-plane contribution. For supported MoS$_2$, vacancies are not the dominant dampening factor to the out-of-plane motion because the substrate effects are much stronger. As before, in Fig. 4(b) we also show the total thermal conductivity vs. vacancy density based on kinetic theory (dashed line). The trend agrees well with our simulations, suggesting that effects of the substrate and defects on thermal conductivity are not coupled. Overall, even at higher defect densities, the thermal conductivity of supported MoS$_2$ is significantly lower than of suspended MoS$_2$ (for the same defect density), meaning the substrate always plays a substantial role in reducing the thermal conductivity of monolayer MoS$_2$.

**D. Bilayer MoS$_2$**

In addition to 1L MoS$_2$, we also investigate the thermal properties of the bilayer (2L) material, which is of interest for electronics because it has smaller band gap, lower contact resistance, and generally higher mobility [102, 103]. We repeat the previous simulation protocol but with a Bernal-stacked (ABA) 2L MoS$_2$. The resulting *suspended* bilayer thermal conductivity, seen in Fig. 5(a), is $\kappa_{2L} = 94.6 \pm 1.6$ Wm$^{-1}$K$^{-1}$ and is consistent with previous 2L MoS$_2$ simulations [51]. The in-plane contribution is $73.0 \pm 2.1$ Wm$^{-1}$K$^{-1}$ and out-of-plane atomic vibrations contribute $21.6 \pm 0.9$ Wm$^{-1}$K$^{-1}$, representing a ~15% and ~30% decrease from suspended 1L MoS$_2$. For suspended 2L, we note out-of-plane motion contributes a smaller proportion of the thermal conductivity compared to 1L MoS$_2$. Previous studies attributed this to a change in the phonon dispersion as well as an increase in flexural phonon scattering rates [27, 52], The drop in thermal conductivity from 1L to 2L is also consistent with experiment [27], although the measurement uncertainty does not yield a definitive trend.

Our *supported* 2L MoS$_2$ calculations, seen in Fig. 5(b), reveal a thermal conductivity of $46.8 \pm 1.8$ Wm$^{-1}$K$^{-1}$ (~50% decrease vs. suspended 2L) with in-plane and out-of-plane contributions of $38.5 \pm 1.8$

Wm$^{-1}$K$^{-1}$ (~47% decrease) and 8.3 ± 0.3 Wm$^{-1}$K$^{-1}$ (~61% decrease), respectively. This thermal conductivity is 50% *larger* than supported 1L MoS$_2$. Thus, given it has double the thickness, supported 2L MoS$_2$ can carry three times more heat than supported 1L MoS$_2$. The top layer of 2L MoS$_2$ is partly "shielded" (screened) from remote phonon scattering with the a-SiO$_2$, better maintaining intrinsic behavior and yielding a higher thermal conductivity than supported 1L MoS$_2$. Again, we find the suppression of in-plane atomic motion drives the overall reduction in thermal conductivity. Experimentally, a larger thermal conductivity in supported 2L MoS$_2$ than 1L MoS$_2$ has also been observed [57].

These results have interesting implications for 2L-based electronic devices because, in addition to improved lateral heat flow (as seen here), previous work has also suggested that 2L MoS$_2$ has a lower thermal boundary resistance with SiO$_2$ than 1L MoS$_2$ [104], i.e. better cross-plane heat flow. In other words, heat removal from 2L-based MoS$_2$ devices is expected to be better than 1L devices *all-around*. Thus, 2L MoS$_2$ could be more attractive for flexible electronics and integrated circuit applications, where heat removal is more important, in addition to its electronic advantages mentioned earlier.

**E. Encased Monolayer and Bilayer MoS$_2$**

For technological reasons such as encapsulation, doping, or top-side gating, MoS$_2$ devices are often encased by a superstrate, such as an oxide [17, 103, 105, 106]. In order to simulate these circumstances, we duplicate the substrate and place it above MoS$_2$, creating the encased 1L and 2L MoS$_2$ structures shown in the insets of Fig. 6. In Fig. 6(a), we find the thermal conductivity of encased 1L MoS$_2$ is 22.1 ± 1.8 Wm$^{-1}$K$^{-1}$, with in-plane and out-of-plane contributions of 18.3 ± 1.4 Wm$^{-1}$K$^{-1}$ and 3.8 ± 0.6 Wm$^{-1}$K$^{-1}$, respectively. Compared to supported 1L MoS$_2$, the thermal conductivity drops an additional ~28%. A similar degradation of the thermal conductivity of encased graphene was observed experimentally [60], but with a ~70% decrease from supported [7] to encased [60]. However, we note that in the encased experiments [60] the graphene may have been damaged during the top SiO$_2$ layer deposition, partly causing the lower thermal conductivity.

Surprisingly, the 1L MoS$_2$ out-of-plane contribution only dropped by ~17% from supported to encased structures, in stark contrast to the ~85% drop from suspended to supported structures. This suggests a substrate already suppresses most out-of-plane motion, and a superstrate cannot suppress it much further. Additionally, comparisons of the VDOS calculations [as in Fig. 2(d) but for encased MoS$_2$] reveal non-negligible changes in the out-of-plane VDOS for sulfur atoms (~10-15% reduction for superstrate structure) for frequencies above 10 THz. These frequencies are in the optical phonon range which do not contribute significantly to thermal conductivity in MoS$_2$ due to low phonon velocities. The main reduction factor is likely the large number of vibrational modes in a-SiO$_2$ (encased structure has twice as many as supported structure) that MoS$_2$ phonons can interact with.

We also examine encased 2L MoS$_2$ as shown in Fig. 6(b). We repeat the simulation protocol verbatim except we reduce the run time from 10 ns to 5 ns when the thermal conductivity appears sufficiently converged. The thermal conductivity of encased 2L MoS$_2$ is 36 ± 0.2 Wm$^{-1}$K$^{-1}$, a decrease of ~23% from the supported 2L structure (and a decrease of ~62% from suspended bilayer). However, this reduction is proportionally less than that experienced by the encased 1L MoS$_2$. Overall, encased 2L MoS$_2$ has a thermal conductivity ~63% larger than and can carry over three times the heat of encased 1L MoS$_2$. Thus, 2L MoS$_2$ will have a higher thermal conductivity than the monolayer, both when interacting with a substrate and/or superstrate, making 2L MoS$_2$ more attractive for applications with larger heat removal requirements. An increase in thermal conductivity with number of layers was also measured in encased graphene around room temperature [60]. As it did in graphene, we expect the thermal conductivity of encased MoS$_2$ to increase with number of layers up to the bulk MoS$_2$ thermal conductivity value (83 ± 3 Wm$^{-1}$K$^{-1}$ for MoS$_2$ using this potential [51]). Given that remote phonon scattering (with the substrate) only penetrates up to ~1 nm into the MoS$_2$ (see section 3 of the Supplement), we expect the thermal conductivity of encased MoS$_2$ to converge to the bulk value within a few layers. However, an extended study on the layer-dependent thermal conductivity for supported and encased MoS$_2$, up to bulk-like thickness, is left for follow-up work.

## IV. SUMMARY AND CONCLUSIONS

We investigated the effects of an SiO$_2$ substrate and encapsulation on the in-plane thermal conductivity of MoS$_2$ using molecular dynamics simulations. Figure 7 summarizes the thermal conductivities of all structures considered. The thermal conductivity of 1L MoS$_2$ decreases from ~117 Wm$^{-1}$K$^{-1}$ (suspended) to ~31 Wm$^{-1}$K$^{-1}$ when supported by an a-SiO$_2$ substrate, a drop of ~74% due to remote phonon scattering with a-SiO$_2$ vibrational modes. While out-of-plane atomic motion is more sensitive to substrate effects, we found the *dominant* mode of thermal transport drives the overall reduction in thermal conductivity of supported 2D materials; for MoS$_2$ it is the in-plane atomic motion, for graphene it is the out-of-plane atomic motion.

Our simulations suggest that a large range of defect concentrations could explain the range of thermal conductivities measured for suspended MoS$_2$ in the literature. However, the thermal conductivity of supported MoS$_2$ appears less sensitive to sulfur vacancy defects (up to 5×10$^{13}$ cm$^{-2}$) and temperature (up to 700 K) than suspended MoS$_2$. In both supported and encased structures we found 2L has >50% higher thermal conductivity than 1L MoS$_2$, thus it can carry over three times more heat. In other words, for certain applications (like integrated circuits) 2L (or slightly thicker) MoS$_2$ could be preferred from a purely thermal point of view because it suffers less from substrate or encapsulation effects than 1L MoS$_2$. However, thicker films could also have drawbacks from cross-plane heat transport [25], and ultimately applications must consider a combined electro-thermal design. Overall, our results will lead to more informed device or system designs with 2D materials from a thermal perspective.

**Acknowledgements:** We thank Dr. Zheyong Fan for the constructive discussion about modifying the GPUMD code for the purposes of this project. AJG acknowledges support from an NDSEG Fellowship. This work was also supported by ASCENT, one of the six centers in JUMP, a Semiconductor Research Corporation (SRC) program sponsored by DARPA. We also thank Stanford University and the Stanford Research Computing Center (Sherlock cluster) for providing computational resources and support.

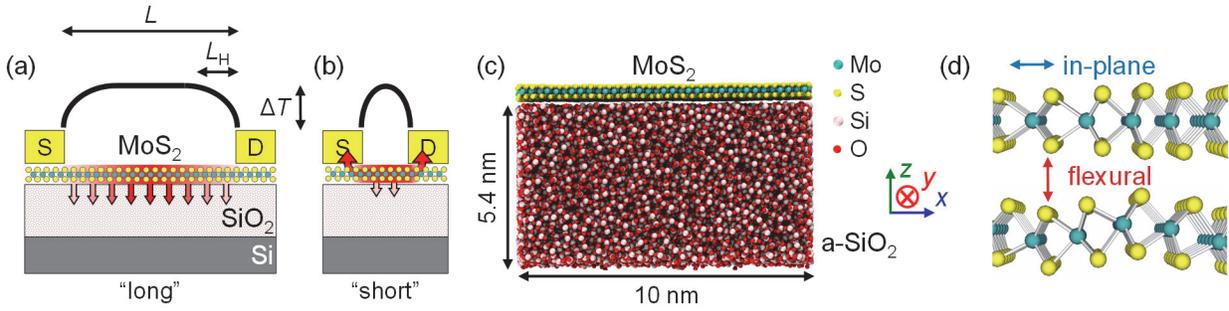

**Figure 1**: Cartoons of thermally (a) "long" and (b) "short" MoS$_2$ devices, i.e. transistors or interconnects. The thermally "long" device dissipates heat mostly through the substrate (see red arrows), whereas the "short" device sinks heat mostly through its contacts. The "thermal length" of the device is with respect to the thermal healing length, $L_H \sim 0.1$ μm for MoS$_2$ on SiO$_2$ substrates. In practice, both device types are often encapsulated by another insulator (not shown, but discussed later). (c) Simulation domain of monolayer MoS$_2$ (10×10 nm$^2$) supported by 5.4 nm of amorphous SiO$_2$. Suspended MoS$_2$ simulations use an identical MoS$_2$ structure without SiO$_2$. (d) Schematic showing in-plane (top) and flexural (bottom) phonons in MoS$_2$. Dominant atomic motion for in-plane phonons is in the $x$ or $y$ direction, and in the out-of-plane, or $z$ direction, for flexural phonons.

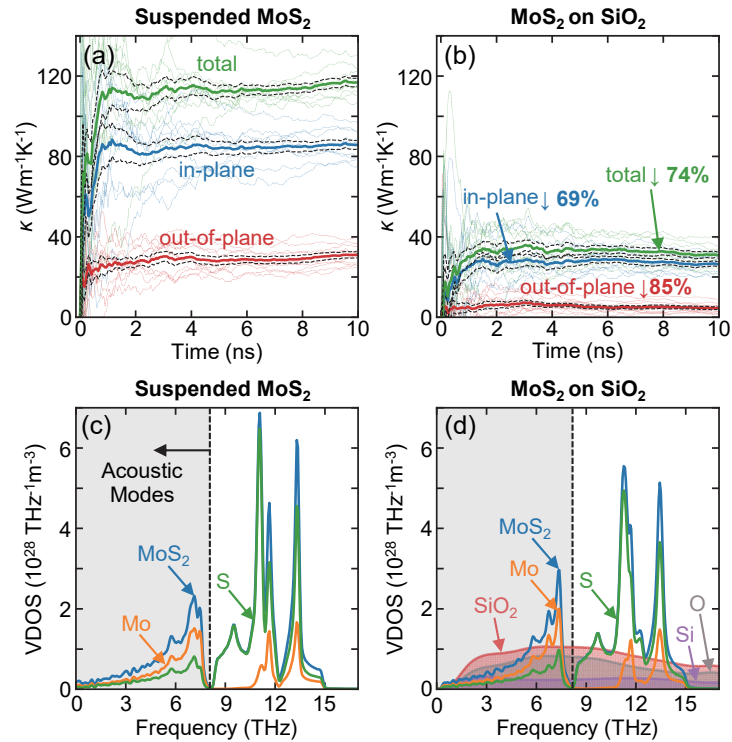

**Figure 2**: Total thermal conductivity (green lines) of (a) suspended 1L MoS$_2$ and (b) SiO$_2$-supported 1L MoS$_2$ including contributions from in-plane (blue lines) and out-of-plane (red lines) atomic motion. Semi-transparent lines represent independent simulations, solid lines represent averages over all runs, and dotted lines show the standard error. The percent reduction in thermal conductivity from suspended to supported MoS$_2$ is labeled in (b). The elemental VDOS for the suspended and supported MoS$_2$ systems are shown in (c) and (d), respectively, with the shaded regions highlighting the heat-carrying, acoustic phonons in MoS$_2$.

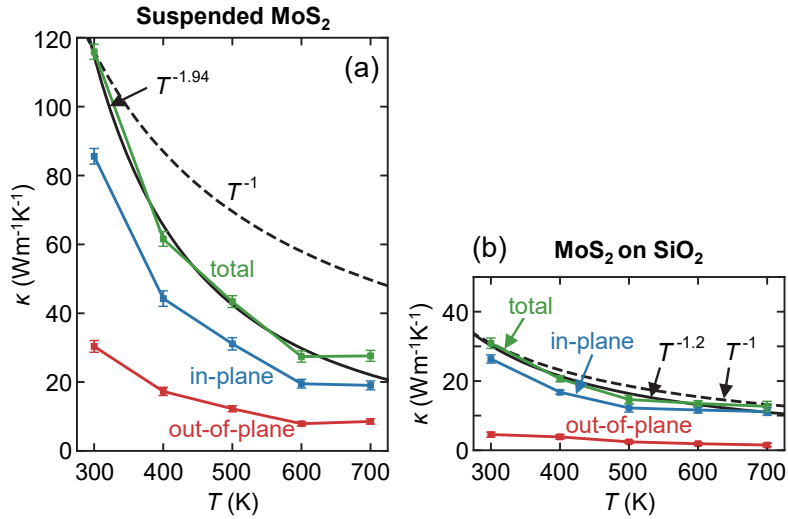

**Figure 3**: Total in-plane thermal conductivity of single-layer MoS$_2$ as a function of temperature for both (a) suspended and (b) supported structures. The dotted black lines illustrate the expected $T^{-1}$ dependence and the solid black lines show a dependence from fits to our calculations.

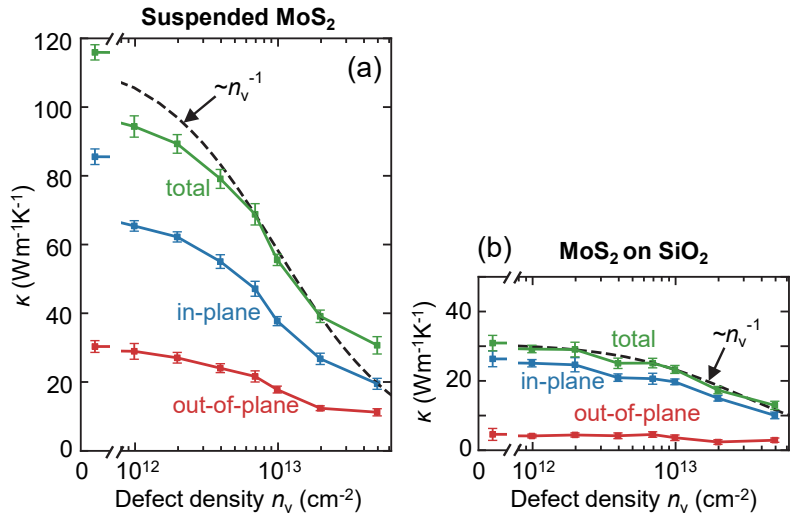

**Figure 4**: (a) and (b) show the in-plane thermal conductivity as a function of defect density for single-layer suspended and supported MoS$_2$. Similarly, the dotted black line illustrates the expected trend with defect density, here fit with $(\eta + \beta n_v)^{-1}$, with further details given in Supplementary section 9.

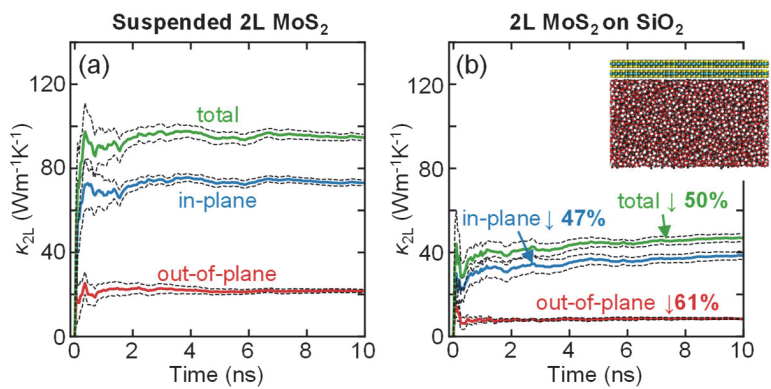

**Figure 5**: Total thermal conductivity (green lines) of (a) suspended 2L $MoS_2$ and (b) supported 2L $MoS_2$ including their in-plane (blue lines) and out-of-plane (red lines) contributions. The percent reduction in thermal conductivity from suspended to supported 2L $MoS_2$ is shown in (b) and the supported 2L $MoS_2$ structure is included in the inset.

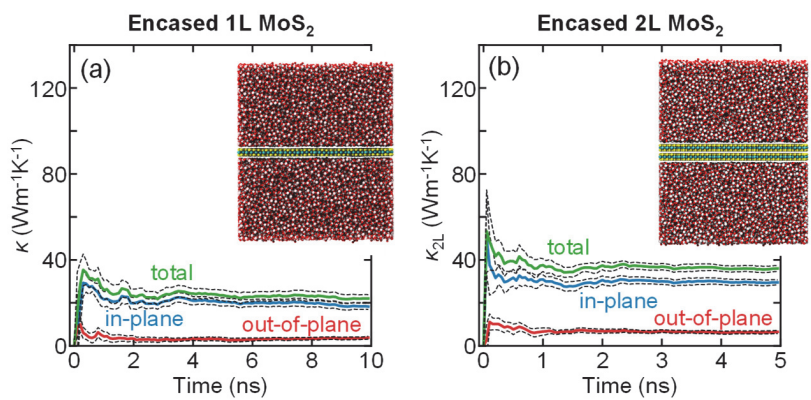

**Figure 6**: Thermal conductivity of encased (a) monolayer and (b) bilayer $MoS_2$ with corresponding structures shown in the insets. Encased 2L $MoS_2$ has a higher thermal conductivity than 1L but both structures have a lower thermal conductivity than their corresponding supported structures.

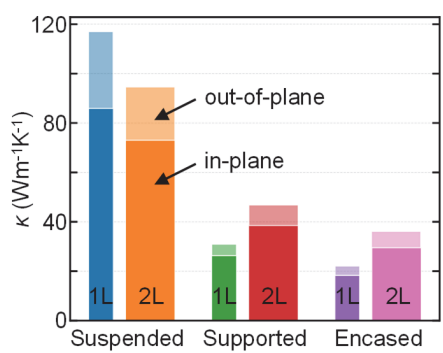

**Figure 7**: Bar chart summarizing the calculations of thermal conductivity for different structures studied in this work. From left to right, the thermal conductivities are: 117.0, 94.6, 30.9, 46.8, 22.1, and 36.0, all in $Wm^{-1}K^{-1}$. The bottom segment of each bar shows the contribution from in-plane atomic motion, and the top segments show that from the out-of-plane atomic motion. The in-plane contribution dominates heat flow in $MoS_2$ and is most strongly affected by the presence or absence of a substrate or encapsulation. (In contrast, the out-of-plane contributions dominate in graphene.)

# Supplementary Information

# Reduced Thermal Conductivity of Supported and Encased Monolayer and Bilayer MoS$_2$

Alexander J. Gabourie[1], Saurabh V. Suryavanshi[1], Amir Barati Farimani[2], and Eric Pop[1,3,4,*]

[1]Department of Electrical Engineering, Stanford University, Stanford, CA 94305, USA
[2]Department of Mechanical Engineering, Carnegie Mellon University, Pittsburgh, PA 15213, USA
[3]Department of Materials Science & Engineering, Stanford University, Stanford, CA 94305, USA
[4]Precourt Institute for Energy, Stanford University, CA 94305, USA
*Contact: *epop@stanford.edu*

## 1. Literature Review of Suspended Monolayer (1L) MoS$_2$ Simulations

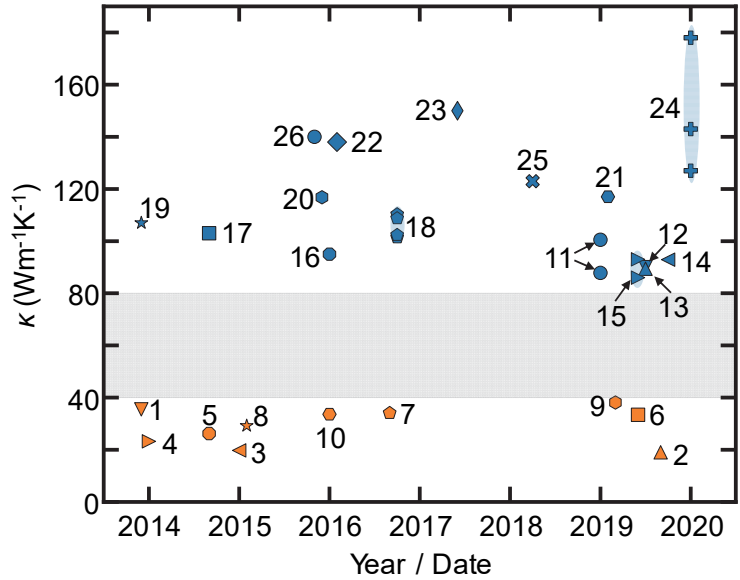

**Figure S1**: Thermal conductivity calculations for room temperature, suspended, monolayer MoS$_2$ vs. publication date. Reported calculations appear to fall in two ranges: the lower range between ~19 to 38 Wm$^{-1}$K$^{-1}$ (orange symbols) [1-10], and the higher range between ~82 to ~178 Wm$^{-1}$K$^{-1}$ (blue symbols) [11-26]. The shaded region shows the range where no reported thermal conductivity values fall.

## 2. Lennard-Jones Parameters

To model van der Waals (vdW) interactions between MoS$_2$ and SiO$_2$ atoms at the interface, we use the Lorentz-Berthelot mixing rules with the appropriate pairwise parameters for each individual atomic element to define our Lennard-Jones (LJ) models. The LJ potential is defined as:

$$V = 4\chi\epsilon\left[\left(\frac{\sigma}{r}\right)^{12} - \left(\frac{\sigma}{r}\right)^{6}\right] \quad (S1)$$

where $\epsilon$ is the potential well depth, $r$ is the distance between particles, $\sigma$ is the distance between particles where $V = 0$, and $\chi$ is a scaling factor for $\epsilon$ sensitivity tests. All calculations in the main text use $\chi = 1$ and the sensitivity of thermal conductivity to $\chi$ is given in section 6. The Lorentz-Berthelot mixing rules define $\epsilon$ and $\sigma$, for atom types A and B, as:

$$\epsilon_{AB} = \sqrt{\epsilon_{AA} \cdot \epsilon_{BB}} \tag{S2}$$

$$\sigma_{AB} = \frac{\sigma_{AA} + \sigma_{BB}}{2}. \tag{S3}$$

The $\epsilon$ and $\sigma$ parameters chosen for Mo and S atoms are from the REBO-LJ potential itself (Mo and S LJ interactions are explicitly defined in this potential for multi-layer simulations) [27, 28], and the Si and O parameters are from the Universal Force Field [29]. The final LJ parameters are in the table below:

**Table S1**: LJ parameters used for pairwise vdW interactions.

| Atomic Pair | $\epsilon$ (meV) | $\sigma$ (Å) |
|---|---|---|
| Mo-Mo* | 0.58595 | 4.20 |
| Mo-S* | 20.0 | 3.13 |
| Mo-Si | 3.1960 | 4.0132 |
| Mo-O | 1.2347 | 3.6591 |
| S-S* | 2.8 | 3.665 |
| S-Si | 18.672 | 3.478 |
| S-O | 7.2137 | 3.4723 |

* Parameters defined explicitly in REBO-LJ potential; not included through separate LJ potential.

### 3. MoS$_2$ Thermal Conductivity Dependence on Amorphous SiO$_2$ (a-SiO$_2$) Thickness

Because amorphous materials have small phonon mean free paths, and because significant interactions between SiO$_2$ and MoS$_2$ do not occur beyond our LJ cutoff distance (1 nm), we do not expect the MoS$_2$ thermal conductivity to depend on the thickness of a-SiO$_2$. We confirm this by computing the thermal conductivity of MoS$_2$ at three different a-SiO$_2$ thicknesses: 2.7 nm, 4.05 nm, and 5.4 nm. The results are shown in Fig. S2 below. The thickness we choose for all simulations in the main text is 5.4 nm.

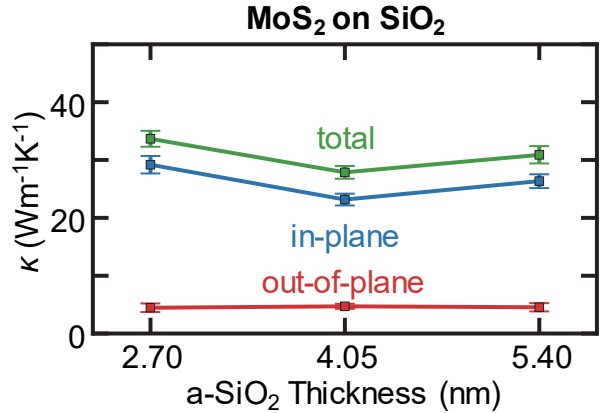

**Figure S2**: The thermal conductivity of supported, monolayer MoS$_2$ vs. a-SiO$_2$ thickness.

### 4. MoS$_2$ on a-SiO$_2$ Structure Creation

The structure we use for supported MoS$_2$ simulations is created and verified in three steps: First, we create the amorphous SiO$_2$ structure. Given the smallest lateral dimensions of MoS$_2$ that reproduced thermal conductivity results [21], we create the thickest SiO$_2$ that can run given the constraints of our compute cluster (Tesla V100-SXM2-16GB GPU, 48 hour time limit). Our final a-SiO$_2$ structure is 10×10 nm$^2$ laterally, 5.4 nm thick and we create it from a crystalline block of 43,200 SiO$_2$ atoms. For the crystalline to amorphous phase transition of SiO$_2$, we run a melt-quench simulated anneal. Using the Munetoh [30] potential, we find very high temperatures are required to melt the crystal structure; we find this to be true for both GPUMD [31, 32] and LAMMPS [33]. We create our final structure using only GPUMD. The simulated anneal conditions are summarized in Fig. S3 below. The entire process is under the constant atom number, volume, and temperature ensemble (*NVT*) as lateral dimensions must remain constant for MoS$_2$ to be unstrained in the final supported structure.

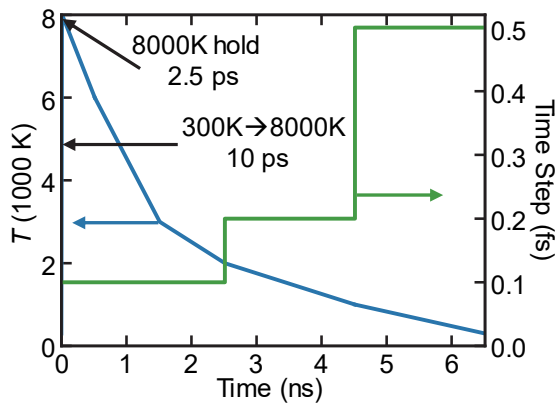

**Figure S3**: Simulated anneal temperature (blue) and time step (green) conditions to create amorphous SiO$_2$.

Second, we introduce a vacuum in the azimuthal direction. This vacuum creates two surfaces on a-SiO$_2$ which may not be stable. It also enables extra stress to be released in the vertical direction, although we do not see a change in SiO$_2$ thickness. We perform a stability check by running three, 250 ps runs: a temperature ramp up to 800 K, an 800 K hold, and a temperature ramp down to 150 K. These runs are also in the *NVT* ensemble with a time step of 0.5 fs. During this process, a few atoms leave the surfaces of SiO$_2$. These atoms are removed and the remaining, stable atoms are left for the final structure. We removed 18 unstable atoms for our final structure leaving a total of 43,182 SiO$_2$ atoms. For all subsequent simulations at all temperatures, no Si or O atoms leave the a-SiO$_2$ block validating the effectiveness of this approach.

The third and final step is to add the MoS$_2$ layer on the a-SiO$_2$ block. The MoS$_2$ sheet is placed ~3 Å (calculated from the bottom S atoms) above the a-SiO$_2$ surface. This configuration is run for 1 ns: the first 500 ps is a temperature ramp from 150 K to 300 K, and the remaining 500 ps is a 300 K temperature hold. The run is in the *NVT* ensemble with a time step of 0.5 fs. During this run we check the stability of the MoS$_2$ on a-SiO$_2$ structure. Our final MoS$_2$ on a-SiO$_2$ structure, shown in Fig. 1(c) of the main text, is taken from the end of this run.

**5. Simulation Protocol**

In all simulations, we use a time step of 0.5 fs as it conserves energy in an ensemble held at constant atom number, volume, and energy (*NVE*). The simulation protocol is as follows: equilibrate the system in the constant atom number, pressure, and temperature ensemble (*NPT*) with zero in-plane pressure and a target temperature of 300 K (unless otherwise specified) for 1 ns. For supported and encased structures, the *NPT* step introduces a ≤ ~|0.7|% compressive strain in MoS$_2$, which we find to marginally reduce its thermal conductivity. After this initialization, we run the system in the *NVT* ensemble using the Nosé-Hoover chain thermostat [34] at the same target temperature for 10 ns. Additionally, a driving force is applied to all atoms for the homogeneous nonequilibrium MD (HNEMD) method, which we use to calculate thermal conductivity. At 10 ns, the thermal conductivity of each structure is sufficiently converged.

## 6. Sensitivity of Thermal Conductivity to $\chi$

Compared to the MoS$_2$ and SiO$_2$ interatomic potentials, which are parameterized based on specific crystal structures and material properties, the LJ interaction parameters are determined by the simplistic Lorentz-Berthelot mixing rules (see section 2). Here, we scale the $\epsilon$ values by $\chi$ to test the sensitivity of the in-plane thermal conductivity of supported 1L MoS$_2$ to this interaction parameter. We perform five simulations per $\chi$ (except for $\chi = 1$, which uses results from the main text) using the same methodology described in section 5 and section II of the main text. The results are below in Fig. S4(a). Unsurprisingly, a weaker interaction strength ($\chi < 1$) leads to a higher thermal conductivity, whereas the opposite happens with a stronger interaction strength ($\chi > 1$). It is important to understand these results in the context of our comparison to suspended MoS$_2$. Figure S4(b) illustrates how the percent reduction in thermal conductivity from suspended MoS$_2$ to supported changes with $\chi$. We see that the range for the reduction is between 65% and 80% over all $\chi$. Ultimately, no matter what the true $\chi$ value is, the supported MoS$_2$ thermal conductivity reduction remains comparable.

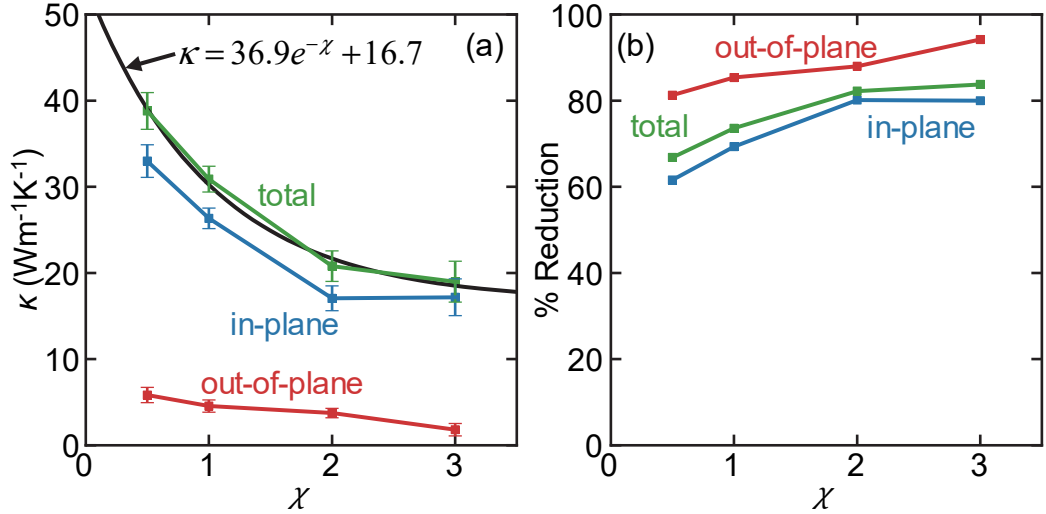

**Figure S4:** (a) Thermal conductivity of supported 1L MoS$_2$ with respect to the LJ scaling factor, $\chi$, which modulates the interaction strength between MoS$_2$ and a-SiO$_2$. The black line shows an analytic fit to the total in-plane thermal conductivity from our simulations. (b) The percent reduction of supported 1L MoS$_2$ thermal conductivity from suspended 1L MoS$_2$ calculations. This highlights how, even over a range of $\chi$, the total in-plane thermal conductivity is still reduced significantly (65% to 80%).

## 7. Vibrational Density of States (VDOS)

The calculations of the VDOS are based on the velocity autocorrelation function (VACF) [35] which states that the Fourier transform (specifically discrete cosine transform) of the VACF can directly give the VDOS. This can be written as:

$$\text{VDOS}_j(\omega) = \int_{-\infty}^{\infty} \cos(\omega t) \text{VACF}_{jj}(t) dt \quad (S4)$$

where the subscript $j$ denotes a Cartesian direction. Here, we define the total VACF as the mass-weighted sum of each atomic VACF [36], which can be written as:

$$\text{VACF}_{jk}(t) = \sum_{i=1}^{N} \left\langle m_i \mathbf{v}_i^j(t) \cdot \mathbf{v}_i^k(0) \right\rangle \quad (S5)$$

where $j$ and $k$ are Cartesian directions, $i$ is the atom index, $m$ and $\mathbf{v}$ are the mass and velocity of atom $i$, and $N$ is the total number of atoms. Since the integration of the VDOS equals the total number of degrees of freedom for a system ($3N$), we normalize the VDOS to meet this criterion. This normalization looks like:

$$\int_0^\infty \frac{\text{VDOS}_j(\omega)}{2\pi} d\omega = 3N . \tag{S6}$$

In Figs. 2(a) and 2(b) in the main text, we also normalize by system volume so $MoS_2$ and $SiO_2$ can be plotted on the same scale. Finally, atom type decompositions of the VDOS simply come from grouping this calculation by atom type. The VDOS calculation was implemented and run in GPUMD.

## 8. Kinetic Theory Approximation - Temperature and Defect Density Dependence

Often used for crystalline materials, the kinetic formula can be used to understand and compute the thermal conductivity. The expression for thermal conductivity can be written as [37]:

$$\kappa = \frac{1}{d} C v \lambda_{ph} \tag{S7}$$

where $d$ is the dimension of the system, $C$ is the heat capacity per unit volume, $v$ is the average particle speed, and $\lambda_{ph}$ is the phonon mean free path. At high temperatures (i.e. above the Debye temperature, $\theta_D$), the heat capacity term transitions to the constant, Dulong-Petit Law value of $C = 3nk_B$ [38], where $n$ is the sample's atom number density, and $k_B$ is the Boltzmann constant. Three-phonon Umklapp scattering is proportional to the temperature, meaning $\lambda_{ph} \propto T^{-1}$. As a result, the thermal conductivity at high temperature ($T > \theta_D$) has the relationship $\kappa \propto T^{-1}$ [37]. However, the rate of decline has been shown to hold a more general trend of $\kappa \propto T^{-x}$ where $1 < x < 2$ [38]. The exponent incorporates the balance between the three- and four-phonon anharmonic scattering terms ($x$ increases with increasing higher order process contribution) [39]. For impure or defective materials, there is an additional scattering term that must be included in $\lambda_{ph}$ along with the phonon-phonon scattering term. The phonon-phonon and defect scattering effects can be combined using Matthiessen's rule. The defect (vacancy) mean free path can be written as $\lambda_v \propto \sim n_v^{-1}$ [37], where $n_v$ is vacancy density.

## 9. Temperature and Defect Density Fitting

Figures 3(a) and 3(b) in the main text each show two different temperature curves from the kinetic theory approximation: $\kappa \propto T^{-1}$ and $\kappa \propto T^{-x}$. This section will describe how these lines are created. For $\kappa \propto T^{-1}$, we plot $\kappa = \alpha T^{-1}$ where $\alpha = \kappa_{MD}(T = 300 \text{ K}) \cdot 300 \text{ K}$ in units of Wm$^{-1}$, with $\kappa_{MD}$ being the molecular dynamics (MD)-calculated thermal conductivity. This means that $\kappa(T = 300K) = \kappa_{MD}(T = 300K)$. This is arbitrarily chosen to aesthetically capture the $T^{-1}$ dependence. All possible choices can be seen in Figs. S5(a) and S5(b), where each line chooses $\alpha$ based on a different temperature between 300 K and 700 K. It also shows that, no matter the choice in $\alpha$, the MD-calculated thermal conductivity of $MoS_2$ decays faster than $T^{-1}$. For $\kappa \propto T^{-x}$, we perform a non-linear, least squares fit using the MD-calculated $MoS_2$ thermal conductivity results. The fit equation was $\kappa = \alpha T^{-x}$, with $\alpha$ and $x$ as fitting parameters. The final parameters of the fit for suspended 1L $MoS_2$ were $\alpha = 7.37 \times 10^6$ Wm$^{-1}$K$^{-(1+x)}$ and $x = 1.9413$ and were $\alpha = 2.78 \times 10^4$ Wm$^{-1}$K$^{-(1+x)}$ and $x = 1.2$ for supported 1L $MoS_2$.

Figures 4(a) and 4(b) in the main text show curves for $\kappa \propto \sim n_v^{-1}$, where $n_v$ is vacancy (defect) density. This curve is based on the vacancy density's effect on the total phonon scattering rate; however, Umklapp scattering must also be accounted for in the calculation of the total phonon scattering rate. The total scattering rate can be written as $\tau^{-1} = \tau_U^{-1} + \tau_v^{-1}$ where $\tau_U^{-1}$ is the Umklapp scattering rate and $\tau_v^{-1}$ is the vacancy scattering rate. Since only $n_v$ is changing, we assume the Umklapp scattering rate is constant. We also know that $\kappa \propto \tau$ (because $\lambda_{ph} = v\tau$) allowing us to write our fitting equation as $\kappa = (\eta + \beta n_v)^{-1}$ with $\eta$ taking the place of the terms related to Umklapp mean scattering time and $\beta$ taking the place of all constants in the vacancy mean scattering time term.

To determine the constant $\eta$, we first consider the case when $n_v = 0$. This gives the direct result of $\kappa(n_v = 0) = \eta^{-1} = \kappa_{MD}(T = 300$ K, $n_v = 0)$ with $\eta$ in units of W$^{-1}$m·K. Solving for $\beta$ we get the expression $\beta = (1-\kappa\eta)\cdot(\kappa n_v)^{-1}$ with $\beta$ in units of m$^3$K·W$^{-1}$. Like the $T^{-1}$ plot, we must choose a parameter, here $n_v$, based on our current available results. In Figs. 4(a) and 4(b) in the main text, we chose $n_v \approx 7\times10^{12}$ cm$^{-2}$ to define our $\beta$ term. Figs. S5(c) and S5(d) show all possible $\kappa \propto \sim n_v^{-1}$ curves from our MD results. We can see that, no matter the choice in $n_v$, the slope is effectively the same meaning our conclusion from the main text remains the same. The decay in thermal conductivity with defect density is what we expect based on kinetic theory.

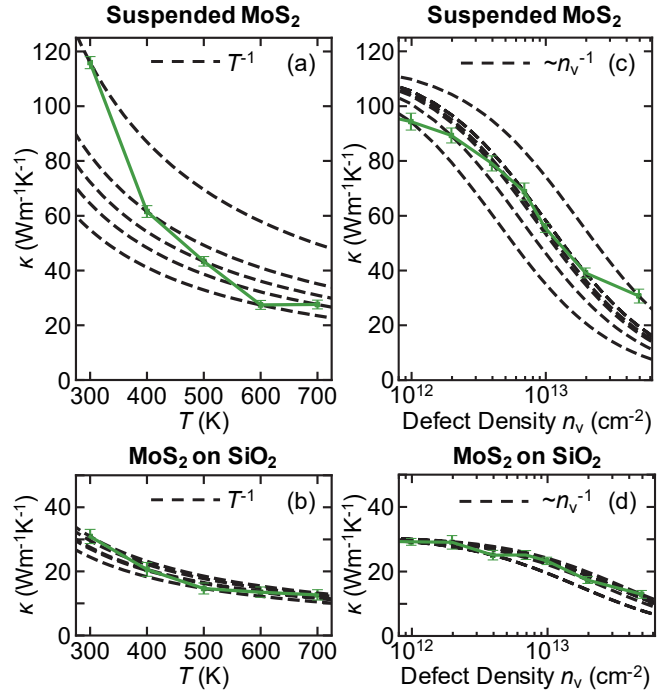

**Figure S5**: Possible fits to kinetic theory curves for (a) suspended and (b) supported 1L MoS$_2$ thermal conductivity vs. temperature as well as fits to kinetic theory curves for (c) suspended and (d) supported 1L MoS$_2$ thermal conductivity vs. sulfur vacancy density. Each curve is fit to a calculated $\kappa_{MD}$ value at a specific temperature or defect density. The plots illustrate how each of the lines look different but the rate of change of thermal conductivity is relatively similar amongst all choices of fitting points.

## 10. Encased Monolayer and Bilayer MoS$_2$ Structures

For encased MoS$_2$ simulations, an additional structure is created. Using the original MoS$_2$ on a-SiO$_2$ structure as a base, we placed a copy of the a-SiO$_2$ block ~3 Å above the MoS$_2$. Using the same check as the final step of section 4, we verify the structure's stability and obtain our final encased monolayer and encased bilayer structures, as seen in Figs. S6(a) and S6(b), respectively. The final monolayer structure has 89,874 total atoms while the final bilayer structure has 93,330 total atoms. These systems required ~3× the computation time per time step as the supported MoS$_2$ structure.

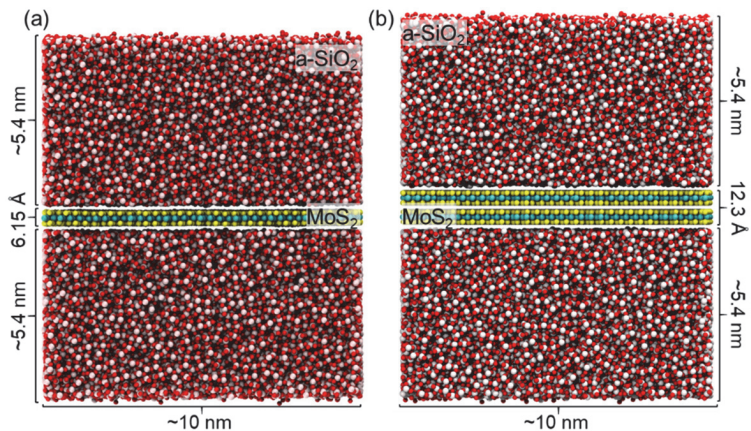

**Figure S6**: SiO$_2$-encased (a) monolayer MoS$_2$ and (b) bilayer MoS$_2$. There are in-plane periodic boundary conditions and there is a vacuum in the out-of-plane direction.

## 11. Amorphous SiO$_2$ – Thermal Conductivity and Vibrational Density of States Benchmarks

Because all supported MoS$_2$ simulations use the same a-SiO$_2$ substrate, we benchmark some of its properties against other studies to determine differences in our systems. First, we calculate the a-SiO$_2$ substrate thermal conductivity. This computation is completed with the Green-Kubo equilibrium MD (EMD) method [40] instead of the HNEMD method. Because the thermal conductivity of a-SiO$_2$ is small (order of ~1 Wm$^{-1}$K$^{-1}$), reliable thermal conductivities can be extracted without a very long EMD run time, and so, instead of tuning the $F_e$ parameter, it is faster to extract thermal conductivity with the EMD method.

The a-SiO$_2$ EMD simulations use a 0.5 fs time step, the same as in the supported MoS$_2$ simulations. Procedurally, we first run in the *NPT* ensemble for 0.25 ns at the desired temperature. Next, we switch to the *NVE* for another 0.25 ns but apply a Langevin thermostat [41] to hold the desired temperature. Under the same conditions, we run an additional 5 ns, this time sampling the heat flux every 5 fs for the EMD method.

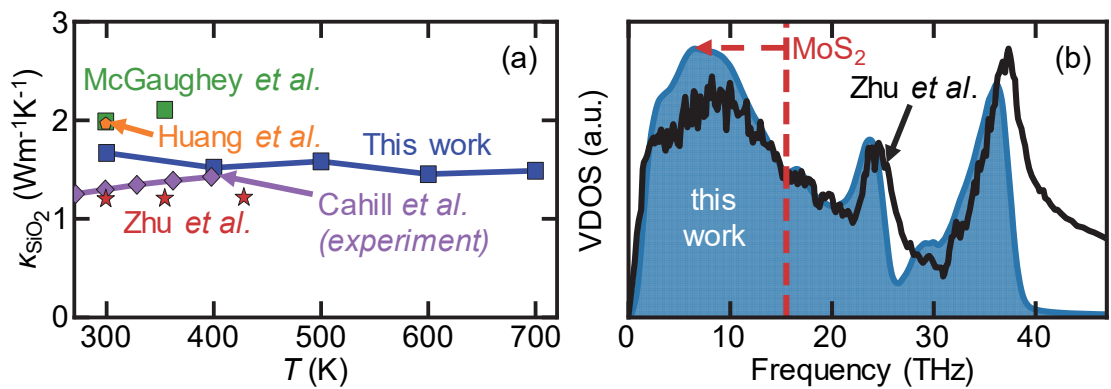

**Figure S7**: (a) Thermal conductivity of a-SiO$_2$ compared to values in literature. (b) Comparison of the VDOS of a-SiO$_2$ with Zhu *et al.*'s work as they used the same Tersoff potential. The curves are normalized by their respective peak VDOS values. The red dotted line denotes highest frequency modes for MoS$_2$.

Our a-SiO$_2$ temperature dependent thermal conductivity, as well as other works in literature, can be seen in Fig. S7(a). Zhu *et al.* [42] uses the same SiO$_2$ Tersoff potential [30] as our work, but we find that our values are ~0.3 Wm$^{-1}$K$^{-1}$ larger than their study. However, we do find that reasonable agreement with Cahill *et al.*'s experimental measurements [43]. From Zhu *et al.*, we note there is no temperature dependence, which agrees with our results. Because the thermal conductivity of amorphous materials tends to increase monotonically with temperature, we expect our overestimation of the thermal conductivity in a-SiO$_2$, compared to experiment, to be less severe at higher temperatures as our calculations show no temperature dependence. We also show MD calculations from McGaughey *et al*. [44] and Huang *et al*. [45] who report MD-calculated thermal conductivities of $\kappa$ = ~2 Wm$^{-1}$K$^{-1}$. Notably, both studies used the BKS potential [46, 47] instead of the Tersoff potential.

We also compare our a-SiO$_2$ VDOS to literature to verify another thermally-relevant property. This behavior is important, as the vibrational states in a-SiO$_2$ will interact with MoS$_2$ and influence our supported MoS$_2$ thermal conductivity calculations. Figure S7(b) shows the total VDOS for our and Zhu's [42] a-SiO$_2$ with each curve normalized by their respective maximums. Qualitatively, these plots are similar at lower frequencies (< 20 THz) but peaks from Zhu's work are shifted to slightly higher frequencies. The discrepancies may come from differences in structure. The red dashed line in Fig. S7(b) denotes the highest frequency at which MoS$_2$ has any phonons. To the left of this line, where most interactions with MoS$_2$ will occur, the VDOS are in reasonable agreement and so our a-SiO$_2$ is well-behaved.